\documentclass{PoS}
\usepackage{epsfig}

\PoS{PoS(LAT2005)243}

\title{QED$_3$ on a space-time lattice: a comparison between compact and 
noncompact formulation }

\ShortTitle{QED$_3$ on a space-time lattice: a comparison between compact and 
noncompact formulation }

\author{Roberto Fiore, \speaker{Pietro Giudice}, Domenico Giuliano,
Donatella Marmottini and Alessandro Papa \\
Universit\`a della Calabria \& INFN Cosenza \\
E-mails: \email{fiore@cs.infn.it, giudice@cs.infn.it, giuliano@cs.infn.it, 
marmotti@cs.infn.it, papa@cs.infn.it}}

\author{Pasquale Sodano \\
        Universit\`a di Perugia \& INFN Perugia \\
        E-mail: \email{sodano@pg.infn.it}}

\abstract{Quantum electrodynamics in a (2+1)-dimensional space-time has been 
object of studies both as effective theory for the pseudogap phase of 
high-T$_c$
superconductors and for the theoretical investigation of mechanisms of
confinement in presence of matter fields.
We discretize the theory using both compact and noncompact formulations,
analyze the behavior of the chiral condensate and of the
monopole density and compare them. Finally we draw some conclusions about the
possible equivalence of the two lattice formulations.}

\FullConference{XXIIIrd International Symposium on Lattice Field Theory\\
25-30 July 2005\\
Trinity College, Dublin, Ireland}

\newcommand{\bi}  {\begin{itemize}}
\newcommand{\ei}  {\end{itemize}}
\newcommand{\be}  {\begin{enumerate}}
\newcommand{\ee}  {\end{enumerate}}

\newcommand{\bc}  {\begin{center}}
\newcommand{\ec}  {\end{center}}

\begin{document}

\section{\label{sec:1} Introduction}

Quantum electrodynamics in 2+1 dimensions (QED$_3$) is an interesting
testfield for understanding the mechanism of confinement in
gauge theories~\cite{Po}, and for the effective description of
low-dimensional, correlated, electronic condensed matter systems,
like spin systems~\cite{us,senthil}, or high-$T_c$ superconductors~\cite{MaPa}. 
The compact formulation of QED$_3$ is more suitable for studying the 
mechanism of confinement, while both compact~\cite{AM} and noncompact formulations 
arise in condensed matter systems. This work is devoted to clarify the relationship 
between these two formulations of QED$_3$ on the lattice (for more 
details, see Ref.~\cite{paper}).

Compact QED$_3$ without fermion degrees of freedom is always confining~\cite{Po};
a charge and anti-charge pair is confined by a linear potential,
as an effect of the proliferation of magnetic monopoles.
If matter fields are introduced, the interaction between monopoles
could turn from $1/x$ to $-\ln{(x)}$ at large distances $x$~\cite{deconf},
so that the deconfined phase may become stable. However,
it has been proposed that compact QED$_3$ with massless fermions is always in
the confined phase~\cite{HeSe,AzLu}.

\begin{figure}[tb]
\bc
\epsfig{file=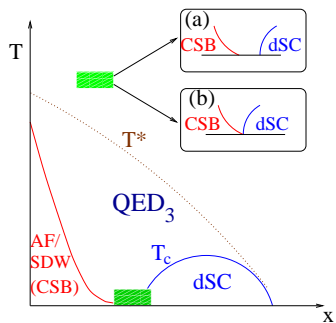,scale=0.35}
\vspace{-2mm}
\caption{Phase diagram in the $(x,T)$ plane~\cite{Fra};
here $x$ represents the doping and $T$ the temperature.}
\label{phase_diagram}
\ec
\end{figure}

At finite $T$ noncompact QED$_3$ is relevant in the
analysis of the pseudo gap phase~\cite{pseudo} of cuprates. 
In Fig.~\ref{phase_diagram} we report the phase diagram of high-$T_c$
cuprates. The small-$x$ phase ($x$ is the doping) is characterized~\cite{HaTh} 
by an insulating antiferromagnet (AF); by increasing $x$, this phase evolves into 
a spin density wave (SDW), that is a weak antiferromagnet. The pseudo gap phase is
located between this phase and the $d$-wave superconducting (dSC) one.
The effective theory of the pseudo gap phase~\cite{pseudo} turns out to be
QED$_3$~\cite{MaPa,Fra,Her}, with spatial anisotropies in the
covariant derivatives~\cite{HaTh}, 
and with Fermionic matter given
by spin-$1/2$ chargeless excitations of the superconducting state
(spinons). These excitations are minimally coupled to a massless
gauge field, which arises from the fluctuating topological defects
in the superconducting phase. The SDW order parameter is the chiral condensate
$\langle \bar{\psi} \psi \rangle$~\cite{Her}. 
If $\langle \bar{\psi} \psi \rangle$ is different from zero, then the $d$-wave
superconducting phase is connected to the spin density wave one 
(see Fig.~\ref{phase_diagram} case b); otherwise the two phases are separated 
at $T=0$ by the pseudo gap phase (see Fig.~\ref{phase_diagram} case a).

The determination of the critical number of flavors, $N_{f,c}$, such that for 
$N_{f} < N_{f,c}$ the chiral condensate is nonzero, is a crucial point 
addressed in many works (see Ref.~\cite{paper} for a detailed list). Here, 
we shall not try to ascertain $N_{f,c}$. However, we will handle the problem of
confinement in presence of massless fermions, by looking at the
relation between monopole density and fermion mass. Moreover we shall
focus on a comparative study of the two lattice formulations of
QED$_3$, the compact and the noncompact ones. In particular, we revisit the 
analysis of Fiebig and Woloshyn of Refs.~\cite{FiWo1,FiWo2}, where the dynamic 
equivalence between the two formulations of (isotropic) QED$_3$ is claimed to be
valid in the finite lattice regime. 

\section{\label{sec:2} Compact and noncompact formulations}

We adopt here the definition of the QED$_3$ parity-conserving continuum Lagrangian 
density in Minkowski metric given in Ref.~\cite{Ap}, 
\begin{equation}
{\cal L} =-\frac{1}{4} F_{\mu \nu}^2+\overline\psi_i iD_\mu \gamma^\mu \psi_i -
 m_0\overline\psi_i \psi_i \;,
\end{equation}
where 
$\psi_i$  ($i=1, \dots, N_f$) are 4-component spinors. 
Since QED$_3$ is a super-renormalizable theory,
dim$[e]=+1/2$, the coupling does not display any energy dependence.
For the definition of the $\gamma$ matrices, for the chiral and parity properties 
of the theory we refer to Ref.~\cite{paper}.

The lattice Euclidean action~\cite{AlFaHaKoMo,HaKoSt} using
staggered fermion fields $\overline\chi,\chi$, is given by
\begin{equation}
\label{action}
 S=S_G+\sum_{i=1}^{N} \sum_{n,m}  \overline\chi_i(n)
M_{n,m} \chi_i(m)\;,
\end{equation}
where $S_G$ is the gauge field action and $M_{n,m}$ is the fermion matrix.
The action (\ref{action}) allows to simulate $N=1,2$ flavours of
staggered fermions corresponding to $N_f=2,4$ flavours of
4-component fermions $\psi$~\cite{BuBu}. 
For the compact formulation $S_G$ is the standard Wilson action
written in terms of the plaquette variable $U_{\mu \nu}(n)$.
Instead, in the noncompact formulation one has
\begin{equation}
S_G[\alpha]= \frac{\beta}{2} \sum_{n,\mu < \nu} F_{\mu \nu}(n)F_{\mu \nu}(n)\;,
\;\;\;
F_{\mu \nu}(n)=  \{ \alpha_{\nu}(n+\hat{\mu})-
  \alpha_{\nu}(n) \} - \{ \alpha_{\mu}(n+\hat{\nu})-
  \alpha_{\mu}(n) \} \;,
\end{equation}
where $\alpha_{\mu}(n)$ is the phase of the link variable $U_{\mu}(n)$ and
$\beta=1/(e^2 a)$.

Monopoles are detected in the lattice using the method given by
DeGrand and Toussaint~\cite{DeTo}.
In the noncompact formulation of QED$_3$ monopoles are not
classical solutions as in compact QED$_3$, but they could give a contribution to the
Feynman path integral owing to the periodic structure of the
fermionic sector~\cite{HaWe}.

\begin{figure}[tb]
\bc
\hspace{-0.5cm}\epsfig{file=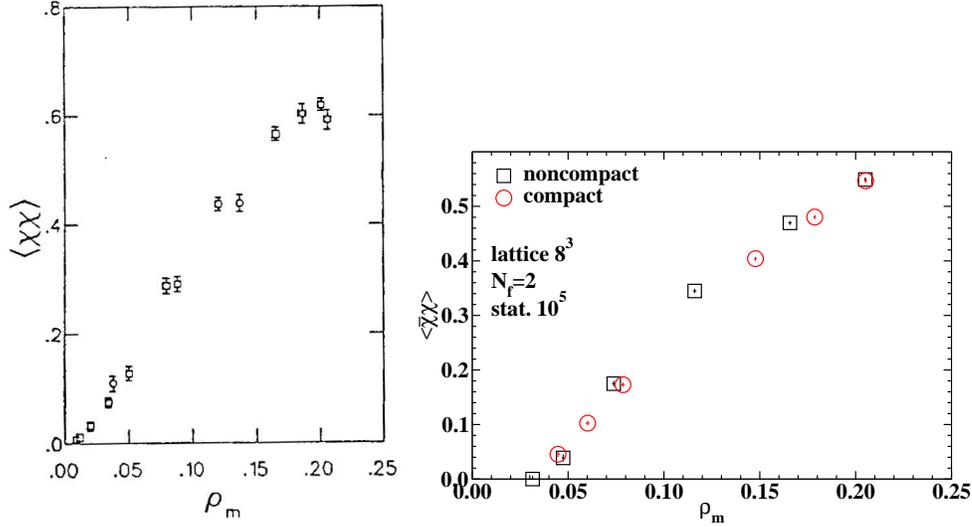,scale=0.25}
\epsfig{file=graf_L8_psi_vs_rho.eps,scale=0.25}
\vspace{-2mm}
\caption[]{Left: Correlation between $\langle\overline{\chi}\chi\rangle$ 
and $\rho_m$ for the compact (circles) and the noncompact (boxes) theories 
for $N_f=2$ and $8^3$ lattice according to Ref.~\cite{FiWo2} (left) and to 
our results (right).}
\label{Fie_Wol}
\ec
\end{figure}

As a signal for continuum physics, we look for plateau of dimensionless
observables, such as $\beta^2 \langle\overline\chi\chi\rangle$.
There are two regimes: for $\beta$ larger than a certain value, the theory is in
the continuum limit, otherwise the system is in a phase with finite
lattice spacing, describing a lattice condensed-matter-like system.
In Refs.~\cite{FiWo1,FiWo2} it is shown there that, when 
$\langle\overline{\chi}\chi\rangle$ is plotted versus the monopole density $\rho_m$, 
data points for QED$_3$ with $N_f=0$ and $N_f=2$ in the compact and noncompact 
formulations in the lattice regime fall on the same curve to a good 
approximation (see Fig.~\ref{Fie_Wol} (left)). This led the
authors of Refs.~\cite{FiWo1,FiWo2} to conclude that the physics of the
chiral symmetry breaking is the same in the two theories.
We want to verify, by the same method, if this conclusion can be extended to 
the continuum limit.

\section{\label{sec:4} Numerical results}

Our Monte Carlo simulation code was based on the hybrid updating algorithm,
with a microcanonical time step set to $dt=0.02$. We simulated one flavour of
staggered fermions corresponding to two flavours of 4-component fermions. Most
simulations were performed on a $12^3$ lattice, for bare quark mass ranging in
the interval $am=0.01\div 0.05$. We made refreshments of the gauge
(pseudofermion) fields every 7 (13) steps of the molecular dynamics. In order
to reduce autocorrelation effects, ``measurements'' were taken every
50 steps. Data were analyzed by the jackknife method combined with binning.

As a first step, we have reproduced the results by Fiebig and Woloshyn which
are shown in Fig.~\ref{Fie_Wol} (left). We find that also in our case data points from
the two formulations nicely overlap (see Fig.~\ref{Fie_Wol} (right)).

\begin{figure}[tb]
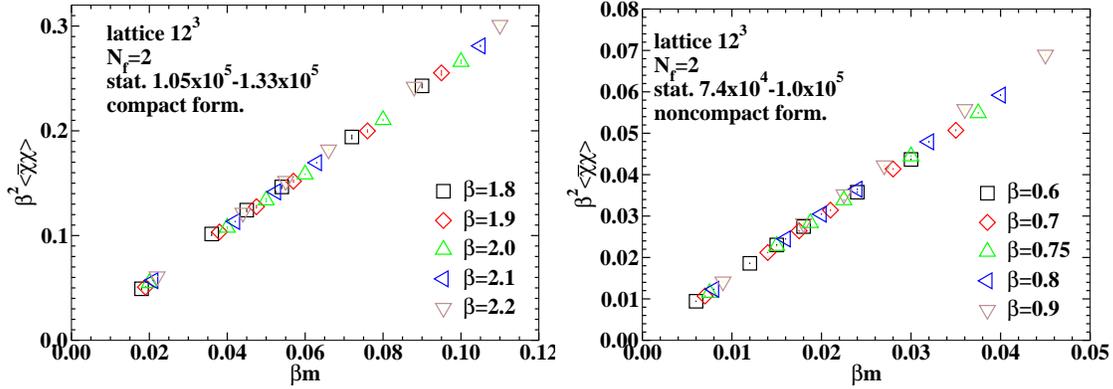

\bc
\epsfig{file=graf_c_beta_2_x_cc_vs_beta_m.eps,scale=0.25}
\epsfig{file=graf_nc_beta_2_x_cc_vs_beta_m.eps,scale=0.25}
\vspace{-2mm}
\caption{$\beta^2 \langle\overline\chi\chi\rangle$ versus $\beta m$
in the compact (left) and in the noncompact (right) formulation.}
\label{cond}
\ec
\end{figure}

Then, in Fig.~\ref{cond} (left) we plot data for $\beta^2
\langle\overline\chi\chi\rangle$ obtained in the compact formulation
versus $\beta m$. We restrict our attention to the subset of $\beta$
values for which data points fall approximately on the same curve (this
indicating the onset of the continuum limit)
which in the present case means $\beta=1.9, 2.0, 2.1$, corresponding
to $L/\beta=6.31,6.00, 5.71$. 
The ratio $L/\beta$ fixes the physical volume that is pratically constant
in the considered range.
A linear fit of these data points 
gives $\chi^2$/d.o.f. $\simeq 8.4$ and the extrapolated value for
$\beta m \rightarrow 0$ turns out to be
$\beta^2\langle\overline\chi\chi\rangle = (1.54 \pm 0.25)\times
10^{-3}$. Restricting the sample to the data at $\beta=2.1$, the
$\chi^2/$d.o.f. lowers to $\simeq 1.3$ and the extrapolated value
becomes $\beta^2\langle\overline\chi\chi\rangle = (0.94 \pm
0.28)\times 10^{-3}$, thus showing that there is a strong
instability in the determination of the chiral limit. If instead a
quadratic fit is used for the points obtained with $\beta=1.9, 2.0,
2.1$, we get $\beta^2\langle\overline\chi\chi\rangle = (0.91 \pm
0.45)\times 10^{-3}$ with
 $\chi^2$/d.o.f. $\simeq 8.7$. Owing to the large uncertainty, this
determination turns out to be compatible with both the previous ones.

In Fig.~\ref{cond} (right) we plot data for $\beta^2 \langle\overline\chi\chi\rangle$
obtained in the noncompact formulation versus $\beta m$. Following the
same strategy outlined before, we restrict our analysis to the data obtained
with  $\beta=0.7,0.75,0.8$, which correspond to $L/\beta= 17.14, 16, 15$.
If we consider a linear fit of these data and extrapolate to $\beta
m \rightarrow 0$, we get $\beta^2\langle\overline\chi\chi\rangle =
(0.45 \pm 0.03)\times 10^{-3}$ with $\chi^2$/d.o.f. $\simeq 17$.
Performing the fit only on the data obtained with $\beta=0.8$, for
which a linear fit gives the best $\chi^2/$d.o.f. value $\simeq 16$,
we obtain the extrapolated value
$\beta^2\langle\overline\chi\chi\rangle = (0.66 \pm 0.07)\times
10^{-3}$. Therefore, also in the noncompact formulation the chiral
extrapolation resulting from a linear fit is largely unstable. A
quadratic fit in this case gives instead a negative value for
$\beta^2\langle\overline\chi\chi\rangle$.

The comparison of the extrapolated value for
$\beta^2\langle\overline\chi\chi\rangle$ in the two formulations is
difficult owing to the instabilities of the fits and to the low
reliability of the linear fits, as suggested by the large values of
the $\chi^2/$d.o.f. Taking an optimistic point of view, one could
say that the extrapolated $\beta^2\langle\overline\chi\chi\rangle$
for $\beta=2.1$ in the compact formulation is compatible with the
extrapolated value obtained in the noncompact formulation for
$\beta=0.8$.

It is worth mentioning that our results in the noncompact formulation
are consistent with known results: indeed, if we carry out a linear fit
of the data for $\beta=0.6,0.7,0.8$ and $am$=0.02, 0.03, 0.04, 0.05 and
extrapolate, we get $\beta^2\langle\overline\chi\chi\rangle =
(1.30 \pm 0.07)\times 10^{-3}$
with an admittedly large $\chi^2/$d.o.f. $\simeq 20$, but very much in
agreement with the value $\beta^2\langle\overline\chi\chi\rangle
= (1.40 \pm 0.16)\times 10^{-3}$ obtained in Ref.~\cite{AlFaHaKoMo}.
We stress that our results are plagued by strong finite volume
effects, therefore our conclusions on the extrapolated values of
$\beta^2\langle\overline\chi\chi\rangle$ are significant only in the
compact versus noncompact comparison we are interested in. 

\begin{figure}[tb]
\bc
\epsfig{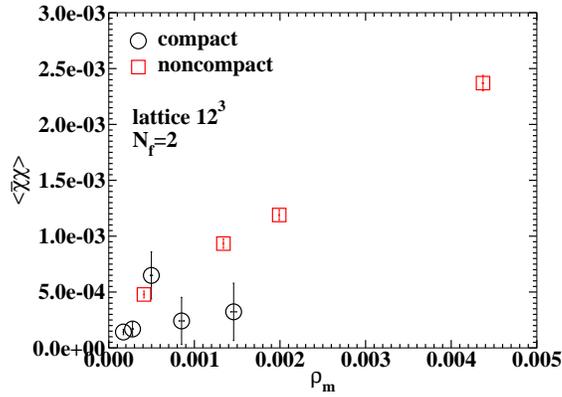}
\vspace{-2mm}
\caption{$\langle\overline\chi\chi\rangle$ versus $\rho_m$ in both
the compact and the noncompact formulations on a $12^3$ lattice.}
\label{cond_vs_rho}
\ec
\end{figure}

In Fig.~\ref{cond_vs_rho} we plot $\langle\overline\chi\chi\rangle$ versus
the monopole density $\rho_m$. Differently from Fig.~\ref{Fie_Wol},
it is not evident with the present results that the two formulations
are equivalent also in the continuum limit, although such an equivalence
cannot yet be excluded. We arrive at the same conclusion considering a 
$32^3$ lattice.

We studied also $\beta^3\rho_m$ versus $\beta m$ for the two formulations
(for plots and more details, see Ref.~\cite{paper}). Our result show that data 
at different $\beta$ values do not fall on a single curve, this suggesting 
that the continuum limit has not been reached for the monopole density
for $\beta=2.2$ in the compact formulation and $\beta=0.9$ in the noncompact one.
We found, however, that the monopole density is independent from the fermion mass. 
Since the mechanism of confinement in the theory with infinitely massive fermions, 
i.e. in the pure gauge theory, is based on monopoles and since the monopole density
is not affected by the fermion mass, we may conjecture that this
same mechanism holds also in the chiral limit.

\end{document}